\documentclass[aps,pra,a4,twocolumn,superscriptaddress,showpacs]{revtex4}
\usepackage{psfig,epsfig}
\usepackage{bm}

\begin{document}

\title{Single-passage read-out of atomic quantum memory }

\author{J. Fiur\'{a}\v{s}ek} 
\affiliation{Department of Optics, Palack\'{y} University, 17. listopadu 50,
77200 Olomouc, Czech Republic}

\author{J. Sherson} 
 \affiliation{QUANTOP, Danish
  National Research Foundation Center for Quantum Optics}
\affiliation{Department of Physics
  and Astronomy, University of Aarhus, DK 8000 Aarhus C, Denmark}
\affiliation{Niels Bohr
  Institute, Copenhagen University, Blegdamsvej 17, 2100 Copenhagen
  \O, Denmark}

\author{T. Opatrn\'{y}}
\affiliation{Department of Theoretical Physics, Palack\'{y} University,
17. listopadu 50, 77200 Olomouc, Czech Republic}

\author{E. S. Polzik} 
\affiliation{QUANTOP, Danish
  National Research Foundation Center for Quantum Optics}
\affiliation{Niels Bohr
  Institute, Copenhagen University, Blegdamsvej 17, 2100 Copenhagen
  \O, Denmark}

\date{\today}

\begin{abstract}
A scheme for retrieving quantum information stored in collective
atomic spin systems onto optical pulses is presented. Two
off-resonant light pulses cross the atomic medium in two orthogonal directions
and are interferometrically recombined in such a way that one of
the outputs carries most of the information stored in the medium.
In contrast to previous schemes our approach requires neither
multiple passes through the medium nor feedback on the light after
passing the sample  which makes the scheme very efficient. The
price for that is some added noise which is however small enough 
for the method to beat the classical limits.

\end{abstract}


\maketitle

\section{Introduction}

The development of complex quantum communication networks requires the
interface between light and atomic carriers of quantum information. The photons
are ideal  for transmitting the quantum information over long distances. The
atoms,  on the other hand, are optimally suited for local storage of quantum
states  in the internal ground-state coherences which can exhibit very long
life-time.  Ideally, the atoms should form a \emph{quantum memory} for light
where  the quantum state of the optical beam can be stored, possibly
processed,  and retrieved later on in a controlled manner. The quantum memory
is an essential ingredient of quantum repeaters \cite{Briegel98,Duan01}  that
allow for long-distance quantum communication over realistic noisy quantum
channels and it is also necessary for scalable all-optical quantum computing as
proposed by Knill, Laflamme and Milburn \cite{Knill01}.

Several methods for exchanging the quantum states of light and atoms have been
considered in the literature such as storing the state of a single photon in a
single atom \cite{Cirac97} or coupling the light to an ensemble of atoms
\cite{Kuzmich97,Hald99,Duan01,Kuzmich00,Kozhekin00,Julsgaard04,Kuzmich04,Dantan05}.  
The latter approach seems to be particularly advantageous because 
it is not necessary to employ a cavity and work in the strong coupling regime. 
Instead sufficiently strong interaction is achieved due to the 
collective enhancement effect \cite{Julsgaard01,Geremia04}. An
alternative to direct light-matter state exchange is to first prepare an
entangled state of atoms and light and then teleport the quantum state of light
onto atoms. The entanglement between a photon and a single trapped ion has been
reported \cite{Blinov04} and also entanglement between a photon and an
ensemble of atoms has been observed \cite{Kuzmich04}.  In related experiments,
correlated photon pairs emitted by an atomic ensemble were detected
\cite{Wal03,Kuzmich03,Chou04}.

The storage of a quantum state of light mode in the ensemble of atoms
has been recently demonstrated experimentally \cite{Julsgaard04} by
exploiting the off-resonant interaction of light with a cloud of
cesium atoms held in a glass cell at room temperature
\cite{Kuzmich98,Kuzmich00a,Duan00,Julsgaard01,Schori02}.  In
that experiment, weak coherent states were stored for a few
milliseconds and the fidelity of the stored states exceeded the best
fidelity achievable by a classical measure-and-prepare procedure. The
storage protocol consisted of sending the light beam through the
atomic ensemble, performing a suitable measurement on the outgoing
light and applying an appropriate feedback on atoms with the help of a
magnetic field.  This procedure, however, cannot be directly inverted
to retrieve the quantum state from atoms back onto light due to the
current technical limitations.  The light pulses used in the experiment need
to be about 1 ms long (i.e. $300$ km) and the memory readout would
require storing somewhere the first pulse which interacted with the
atomic ensemble while a second auxiliary light beam is used to measure
the atoms.  The required long duration of the pulses also prevents one
to employ measurement-free retrieval methods based on multiple
passages of the light pulse through the atomic sample
\cite{Kuzmich03b,Fiurasek03,Hammerer04}.

A possible solution to these problems has been suggested by using  two- and
four-pass protocols with the light beam crossing the medium  simultaneously
from two perpendicular directions \cite{Sherson05}. Although high fidelity
readout can theoretically be achieved in these protocols they each suffer from
deficiencies which can decrease their general applicability. The two-pass
protocol is asymmetric with respect to the two conjugate quadratures 
and a unity gain readout is difficult to achieve since it would require
a complicated temporal profile of the coupling light beam. This makes the two-pass
protocol unsuitable for readout of general input states. With the four-pass 
protocol perfect readout can be achieved in principle, however, since
it involves so many passages even a small light reflection loss at each
crossing can seriously decrease the achievable fidelities. This point will be
investigated quantitatively for our protocol below.

\begin{figure}[!b!]
\centerline{\epsfig{file=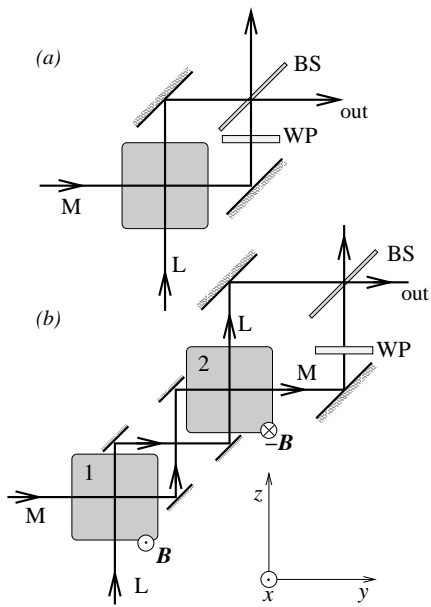,scale=0.3}}
\caption{Scheme for the single-passage read-out of the atomic quantum memory.
  Two light beams L and M simultaneously pass through one $(a)$ or two
  $(b)$ atomic ensembles.  The wave plate (WP) swaps the quadratures
  $x$ and $p$ in one of the beams.  Afterwords, the beams are
  (re)combined on a balanced beam splitter BS. 
\label{fig-scheme}}
\end{figure}

In this paper we suggest an alternative protocol for read-out of the
atomic quantum memory which requires only a single passage of the
light through the atoms and is therefore rather robust with respect to
losses on the glass cell walls. The simultaneous transfer of both
atomic quadratures $x_A$ and $p_A$ onto light is achieved by sending
two light beams perpendicularly to each other through the memory unit,
see Fig. \ref{fig-scheme}.  The scheme can be explained on the
simplified model with a single atomic ensemble as in Fig.
\ref{fig-scheme}a, however, to avoid technical noises, experiments
are usually done using two cells in external magnetic
field (see Fig.  \ref{fig-scheme}b), observing the phenomena on the
sidebands of the
light pulses. As it turns out, these two approaches yield similar
results, however, the number of crosses of the beams through the cell
walls plays a crucial role in increasing the noise.

The rest of the paper is organized as follows. In Section II we derive 
the formula for the output light quadratures and the readout fidelity. 
In Section III we take into account the
effect of losses and show how they influence the read-out fidelity in both the 
single-cell and double-cell schemes. We will also demonstrate that the 
readout can be improved by squeezing the input light beams 
entering the cell. Finally, Section IV contains brief conclusions.

\section{Readout protocol}


\subsection{Single-cell scheme }

Let us first deal with the single-cell scheme with no magnetic field.  
The beam L propagates
along the $z$ axis and the Hamiltonian describing its coupling with
the atomic ensemble is $H_L=a S^L_3 J_z$ and the second beam M
propagates along the $y$ axis and interacts with the ensemble
according to $H_M=a S^M_3J_{y}$ so that the complete Hamiltonian is
\begin{eqnarray}
 H= a \left( S^L_3 J_z + S^M_3J_{y}\right).
\end{eqnarray}
Here
$\bm{S}$ denotes the Stokes vector characterizing the polarization
properties of a light beam, $\bm{J}$ stands for the collective
atomic spin operator and $a$ is the coupling constant \cite{Kuzmich98,Kuzmich00,Duan00}.  
The orientation of the Stokes vectors is chosen such that $S_3$
describes the helicity and $S_1$ describes linear polarization, with
positive $S_1$ corresponding to $x$-polarized light.
The
components of $\bm{S}$ and $\bm{J}$ satisfy the angular momentum-type
commutation relations, $[J_j,J_k]=i\epsilon_{jkl}J_l$
with $j,k,l\in\{x,y,z\}$
and
$[S^L_{j}(t),S^L_{k}(t^\prime)]= i\epsilon_{jkl}\delta(t-t^\prime)
S^L_{l}(t)$, and
$[S^M_{j}(t),S^M_{k}(t^\prime)]=i\epsilon_{jkl}\delta(t-t^\prime)
S^M_{l}(t)$ with $j,k,l\in\{1,2,3\}$.

Both beams M and L
contain strong $x$ polarized component such that the 
$S_1$ components of the Stokes
vectors $S^L_1(t)$ and $S^M_1(t)$ can be approximated by c-numbers. 
We assume that the $S_1$ components are constant throughout the duration 
of the pulses and both beams have the same intensity,
\begin{equation}
S^L_1(t)=S^M_1(t)= \frac{N_L}{2T} \equiv S_1,
\end{equation}
where $N_L$ is the total number of photons in the strong  
component of each beam and $T$ 
is the duration of the pulse. Similarly, the atoms are polarized along the $x$
axis and the $x$ component of the collective atomic spin attains a macroscopic
value and the operator $J_x$ can be replaced with a c-number, $J_x \approx
\langle J_x\rangle$.

In the Heisenberg picture, the atomic quadratures evolve according to
\begin{equation}
\frac{dJ_{y}}{dt}= a\langle J_{x}\rangle S^L_3(t), \qquad 
\frac{dJ_{z}}{dt}= -a\langle J_{x}\rangle S^M_3(t).
\label{HeisenbergJ}
\end{equation}
The relations between input and output $y$ and $z$ components of the Stokes vectors of
beams L and M read \cite{Schori02},
\begin{eqnarray}
S_2^{L, {\rm out}}(t)&=&S_2^{L, \rm in}(t)+aS_1 J_z(t), \nonumber  \\
S_3^{L, \rm out}(t)&=&S_3^{L, \rm in}(t), \nonumber \\
S_2^{M, \rm out}(t)&=&S_2^{M, \rm in}(t)+aS_1 J_y(t),\nonumber  \\
S_3^{M, \rm out}(t)&=&S_3^{M, \rm in}(t).
\label{Sout}
\end{eqnarray}
The Heisenberg equations (\ref{HeisenbergJ}) can be easily solved and yield
\begin{eqnarray}
J_y(t)=J_y(0)+ a\langle J_x\rangle \int_0^t S^L_3(\tau)  d\tau, \nonumber  \\
J_z(t)=J_z(0)- a\langle J_x\rangle \int_0^t S^M_3(\tau)  d\tau.
\label{Jsolution}
\end{eqnarray}

We define the quadratures of the light modes L and M 
in the usual way as integrals over
the whole pulse,
\begin{eqnarray}
x_{L,M} &=& \sqrt{\frac{2}{N_{L}}} \int_0^T S^{L,M}_2(t) dt, \nonumber \\
p_{L,M} &=& \sqrt{\frac{2}{N_{L}}} \int_0^T S^{L,M}_3(t) dt. 
\label{Lquadratures}
\end{eqnarray}
We also define the atomic quadrature operators $x_A=J_y/\sqrt{\langle
  J_x\rangle},$ $p_A=J_z/\sqrt{\langle J_x\rangle}$.  All these
operators satisfy the canonical commutation relations
$[x_j,p_k]=i\delta_{jk}$.  Upon combining the 
equations (\ref{Sout}), (\ref{Jsolution}) and (\ref{Lquadratures})
and changing the order of integration we obtain the expression for
$x_L^{\rm out}$,
\begin{equation}
x_{L}^{\rm out}=x_{L}^{\rm in} + \kappa p_A^{\rm in}-\kappa^2
\int_0^T d t \left(1-\frac{t}{T}\right)\frac{S^M_3(t)}{\sqrt{N_L/2}},
\end{equation}
where we  introduced the effective coupling constant 
$\kappa=a\sqrt{N_L \langle J_x\rangle /2}$.
We can see that the description in terms of the light modes 
(\ref{Lquadratures}) is incomplete
because a mode with non-constant temporal profile $1-t/T$ 
couples to $x_L^{\rm out}$.
The two mode profiles we are dealing with are
\begin{equation}
f_1(t)=1, \qquad f_2(t)= \sqrt{3}\left(1-\frac{t}{T}\right),
\label{f12}
\end{equation}
with the normalization $\int_0^T |f_j^2(t)| dt=T$. 
The two mode functions (\ref{f12}) are not
orthogonal since  $\int_0^T  f_1(t) f_2(t) dt = T \sqrt{3}/2$. 
We can decompose $f_2(t)$
as follows,
\begin{equation}
f_2(t)=\frac{\sqrt{3}}{2}f_1(t)+ \frac{1}{2} f_3(t),
\end{equation}
where $f_3(t)=\sqrt{3}(1-2t/T)$ and the functions 
$f_1(t)$ and $f_3(t)$ are now
orthogonal.We define quadrature operators corresponding to 
the modal profile $f_3(t)$,
\begin{eqnarray}
\tilde{x}_{L,M} &=& \sqrt{\frac{6}{N_L}} \int_0^T \left(1-\frac{2t}{T}\right) 
S^{L,M}_2(t) dt, 
\nonumber  \\
\tilde{p}_{L,M} &=& \sqrt{\frac{6}{N_L}} \int_0^T \left(1-\frac{2t}{T}\right) 
S^{L,M}_3(t) dt.
\label{tildequadratures}
\end{eqnarray}
One can easily check that the quadratures (\ref{tildequadratures}) satisfy 
canonical commutation relations.

Using the above definitions we can express the output quadrature operator
$x_L^{\mathrm{out}}$ as a linear combination of the input light and atomic
quadrature operators. This procedure can be repeated for other quadratures
and after some algebra we finally get:
\begin{eqnarray}
x_L^{\rm out}&=&x_{L}^{\rm in} + \kappa p_A^{\rm in} 
-\frac{\kappa^2}{2}p_M^{\rm in}- \frac{\kappa^2}{2\sqrt{3}} \tilde{p}_{M}^{\rm in}, \nonumber \\
p_L^{\rm out}&=&p_{L}^{\rm in}, \nonumber \\
x_M^{\rm out}&=&x_{M}^{\rm in} + \kappa x_A^{\rm in} 
+\frac{\kappa^2}{2}p_L^{\rm in}+ \frac{\kappa^2}{2\sqrt{3}} \tilde{p}_{L}^{\rm in}, \nonumber \\
p_M^{\rm out}&=&p_{M}^{\rm in}, \nonumber \\
x_A^{\rm out}&=& x_A^{\rm in}+\kappa p_L^{\rm in}, \nonumber \\
p_A^{\rm out}&=& p_A^{\rm in}-\kappa p_M^{\rm in}. 
\label{outquadratures}
\end{eqnarray}

Note that the information about the $x_A$ and $p_A$ quadratures is
stored in two different modes L and M. In order to obtain a single
mode that would contain an approximate version of the state stored
originally in the atomic memory, we have to combine the modes L and M.
To accomplish this, we suggest placing a wave plate
to one of the beams, say M, to switch the quadratures
$x_M\to p_M$, $p_M \to -x_M$,
and to
let these modes interfere on a balanced beam splitter, as shown in
Fig.  \ref{fig-scheme}.  In one of the output arms we
finally obtain a mode with quadratures
\begin{equation}
x=\frac{1}{\sqrt{2}}(x_L^{\rm out}+p_{M}^{\rm out}), \qquad
p=\frac{1}{\sqrt{2}}(p_L^{\rm out}-x_{M}^{\rm out}).
\label{output}
\end{equation}
On inserting the formulas (\ref{outquadratures}) into (\ref{output}), we find
\begin{eqnarray}
x=\frac{\kappa}{\sqrt{2}}p_A^{\rm in} +\frac{1}{\sqrt{2}} x_L^{\rm in}
-\frac{1}{\sqrt{2}}(\frac{\kappa^2}{2}-1) p_M^{\rm in}-\frac{\kappa^2}
{2\sqrt{6}} \tilde{p}_M^{\rm in},
\nonumber \\
p=-\frac{\kappa}{\sqrt{2}}x_A^{\rm in} -\frac{1}{\sqrt{2}} x_M^{\rm in}
-\frac{1}{\sqrt{2}}(\frac{\kappa^2}{2}-1) p_L^{\rm in}-\frac{\kappa^2}
{2\sqrt{6}}
\tilde{p}_L^{\rm in}.
\nonumber  \\
\label{output-kappa}
\end{eqnarray}
If we choose $\kappa=\sqrt{2}$, then the quadratures are transfered from the
atomic memory to the light mode with unit gain,
\begin{eqnarray}
x&=&p_A^{\rm in} +\frac{1}{\sqrt{2}} x_L^{\rm in}-\frac{1}{\sqrt{6}} 
\tilde{p}_M^{\rm in},
\nonumber  \\
p&=&-x_A^{\rm in} -\frac{1}{\sqrt{2}} x_M^{\rm in}
-\frac{1}{\sqrt{6}} \tilde{p}_L^{\rm in}.
\label{finaloutput}
\end{eqnarray}
This is the main result of this paper.

The memory read-out is not perfect due to the noise of the light
quadrature operators in Eq. (\ref{finaloutput}). To be specific, let
us consider a readout of a coherent state $|\alpha\rangle$ stored in
the atomic ensemble. If the light modes are initially in vacuum  states, 
then the added  noise is the same for both
quadratures and can be expressed in terms of the mean photon number
$\bar{n}$ of thermal light with the same quadrature fluctuations
\begin{equation}
\bar{n}=\frac{1}{2}\times\left(\frac{1}{2}+\frac{1}{6}\right)=\frac{1}{3}.
\end{equation}
The fidelity of the memory read-out is  in this case 
\begin{equation}
\mathcal{F}=\frac{1}{1+\bar{n}}
\label{Fidelitynbar}
\end{equation}
and for $\bar{n}=1/3$ we get $\mathcal{F}=3/4$. This value significantly
exceeds the best fidelity 
$\mathcal{F}_{\rm meas}=1/2$ achievable by measure and prepare
procedures, which in \cite{Hammerer05} was shown to be the optimal
classical storage or retrieval procedure.

Note that the readout of the memory is imperfect because a part of the information on the 
state remains in the  memory. In particular, for $\kappa=\sqrt{2}$ the final 
atomic quadratures read
\begin{equation}
x_{A}^{\rm out}=x_A^{\rm in}+\sqrt{2} p_L^{\rm in}, \qquad
p_{A}^{\rm out}=p_A^{\rm in}-\sqrt{2} p_M^{\rm in}.
\label{finalmemory}
\end{equation}
The distribution of the quantum information between the quantum memory and the output 
light beam suggests that the readout protocol is related to the (asymmetric) quantum cloning.
Indeed, in the limit of infinitely squeezed quadratures $\tilde{p}_L=\tilde{p}_M=0$,
the formulas (\ref{finaloutput}) and (\ref{finalmemory}) describe exactly the two clones produced by the optimal Gaussian
asymmetric $1 \rightarrow 2$ cloning machine for coherent states \cite{Fiurasek01}. 
The anti-clone (or, more precisely, the complex conjugated clone) is carried by the 
light beam in the auxiliary output port of the beam splitter, whose quadrature operators 
can be expressed as ($\kappa=\sqrt{2}$),
\begin{eqnarray}
x^{\rm aux}&=&p_A^{\rm in} +\frac{1}{\sqrt{2}} x_L^{\rm in}
-\sqrt{2}p_M^{\rm in}-\frac{1}{\sqrt{6}} \tilde{p}_M^{\rm in},
\nonumber  \\
p^{\rm aux}&=&x_A^{\rm in} +\frac{1}{\sqrt{2}} x_M^{\rm in}
+\sqrt{2}p_L^{\rm in}+\frac{1}{\sqrt{6}} \tilde{p}_L^{\rm in}.
\label{finaloutputaux}
\end{eqnarray}

 
\subsection{Two-cell scheme}
  
To avoid technical noises, one can work with a
two-cell scheme as shown in Fig. \ref{fig-scheme}b \cite{Julsgaard04}.
In this model we assume magnetic field $B$ perpendicular to the two
beams such that the fields in the two cells have opposite orientation.
The atomic spins then rotate around the $x$-axis with the Larmor
frequency $\Omega\propto B$, the direction of rotation being opposite
in the two samples. The atoms interact with light on the sidebands
with frequencies $\omega_0 \pm \Omega$, where $\omega_0$ is the
carrier frequency. The  quadratures of cosine and sine modes 
at sideband frequency $\Omega \gg 1/T$ are defined as follows,
\begin{eqnarray}
 x^{C}_{L,M} &=& \frac{2}{\sqrt{N_L}}\int_0^T S_2^{L,M}(t)\cos(\Omega t) dt,
  \nonumber \\ 
 p^{C}_{L,M} &=& \frac{2}{\sqrt{N_L}}\int_0^T S_3^{L,M}(t)\cos(\Omega t) dt,
  \nonumber \\
 x^{S}_{L,M} &=& \frac{2}{\sqrt{N_L}}\int_0^T S_2^{L,M}(t)\sin(\Omega t) dt,
  \nonumber \\ 
 p^{S}_{L,M} &=& \frac{2}{\sqrt{N_L}}\int_0^T S_3^{L,M}(t)\sin(\Omega t) dt.
  \nonumber 
\end{eqnarray}
The quadratures $\tilde{x}_{L,M}^{C,S}$ and $\tilde{p}_{L,M}^{C,S}$
corresponding to modes with temporal profile $\sqrt{3}(1-2t/T)$ can be defined in
a similar way. 
The interaction  between light and atomic sample 1
can be described by means of the Hamiltonian
\begin{equation}
 H^{(1)}=a(S_3^LJ_z^{(1)}+S_3^MJ_y^{(1)})-\Omega J_x^{(1)},
\end{equation}
where the term proportional to $\Omega$ arises due to the applied magnetic field.
The interaction between light and atomic sample 2
is then described by
\begin{equation}
 H^{(2)}=a(S_3^LJ_z^{(2)}+S_3^MJ_y^{(2)})
 +\Omega J_x^{(2)}.
\end{equation}
The information is encoded in the sum- or difference-quadratures
of the collective atomic spins 
\begin{eqnarray}
 x_{A\pm} &=& \frac{1}{\sqrt{2}}\left( x_A^{(1)}\pm x_A^{(2)}\right)  \\
 p_{A\pm} &=& \frac{1}{\sqrt{2}}\left( p_A^{(1)}\pm p_A^{(2)}\right).
\end{eqnarray} 

The input-output relations can be derived in a similar manner as for
the single-cell scheme by solving the linear Heisenberg equations of motion.
In the limit $\Omega T \gg 1$ which is satisfied in the experiments ($\Omega T
\approx 300$) one thus gets
\begin{eqnarray}
x_L^{C, \rm out} &=& x_L^{C, \rm in} + \kappa p_{A+}^{\rm in} 
- \frac{\kappa^2}{2}  p_M^{C,{\rm in}} - \frac{\kappa^2}{2\sqrt{3}}
  \tilde p_M^{C,{\rm in}} , \nonumber \\
p_L^{C, \rm out} &=& p_L^{C, \rm in} ,  \nonumber \\  
x_M^{C, \rm out} &=& x_M^{C, \rm in} + \kappa x_{A+}^{\rm in} 
 + \frac{\kappa^2}{2}  p_L^{C,{\rm in}} + \frac{\kappa^2}{2\sqrt{3}}
  \tilde p_L^{C,{\rm in}} , \nonumber \\
p_M^{C, \rm out} &=& p_M^{C, \rm in} , \nonumber \\  
x_L^{S, \rm out} &=& x_L^{S, \rm in} - \kappa x_{A-}^{\rm in} 
- \frac{\kappa^2}{2}  p_M^{S,{\rm in}} - \frac{\kappa^2}{2\sqrt{3}}
  \tilde p_M^{S,{\rm in}} , \nonumber \\  
p_L^{S, \rm out} &=& p_L^{S, \rm in} , \nonumber \\  
x_M^{S, \rm out} &=& x_M^{S, \rm in} + \kappa p_{A-}^{\rm in} 
 + \frac{\kappa^2}{2}  p_L^{S,{\rm in}} + \frac{\kappa^2}{2\sqrt{3}}
  \tilde p_L^{S,{\rm in}},  \nonumber \\
p_M^{S, \rm out} &=& p_M^{S, \rm in} ,  
\end{eqnarray}
for the light beams, and for the atomic samples
\begin{eqnarray}
 x_{A+}^{\rm out} &=& x_{A+}^{\rm in} + \kappa p_L^{C,{\rm in}}, \nonumber \\
 p_{A+}^{\rm out} &=& p_{A+}^{\rm in} - \kappa p_M^{C,{\rm in}}, \nonumber \\
 x_{A-}^{\rm out} &=& x_{A-}^{\rm in} + \kappa p_M^{S,{\rm in}}, \nonumber \\
 p_{A-}^{\rm out} &=& p_{A-}^{\rm in} + \kappa p_L^{S,{\rm in}}. 
\end{eqnarray}
So as to have the atomic quadratures in one beam, we place a quarter-wave plate
in beam M, thus switching the quadratures $x_{M}^{C,S} \to p_{M}^{C,S}$
and $p_{M}^{C,S} \to -x_{M}^{C,S}$, and combine the L and M beams on a beam
splitter. In one of the outputs we then find the new quadratures
\begin{eqnarray}
x^{C} &=& \frac{\kappa}{\sqrt{2}}p_{A+}^{\rm in} 
 + \frac{1}{\sqrt{2}} x_L^{C,{\rm in}} 
 \nonumber \\
 & & - \frac{1}{\sqrt{2}}\left(  
 \frac{\kappa^2}{2} -1 \right) p_M^{C,{\rm in}} - \frac{\kappa^2}{2\sqrt{6}}
  \tilde p_M^{C,{\rm in}} ,  \\
p^{C} &=& - \frac{\kappa}{\sqrt{2}}x_{A+}^{\rm in} 
 - \frac{1}{\sqrt{2}} x_M^{C,{\rm in}} 
 \nonumber \\
 & & - \frac{1}{\sqrt{2}}\left(  
 \frac{\kappa^2}{2} -1 \right) p_L^{C,{\rm in}} - \frac{\kappa^2}{2\sqrt{6}}
  \tilde p_L^{C,{\rm in}} , \\
x^{S} &=& -\frac{\kappa}{\sqrt{2}}x_{A-}^{\rm in} 
 + \frac{1}{\sqrt{2}} x_L^{S,{\rm in}} 
 \nonumber \\
 & & - \frac{1}{\sqrt{2}}\left(  
 \frac{\kappa^2}{2} -1 \right) p_M^{S,{\rm in}} - \frac{\kappa^2}{2\sqrt{6}}
  \tilde p_M^{S,{\rm in}} ,  \\
p^{S} &=& - \frac{\kappa}{\sqrt{2}}p_{A-}^{\rm in} 
 - \frac{1}{\sqrt{2}} x_M^{S,{\rm in}} 
 \nonumber \\
 & & - \frac{1}{\sqrt{2}}\left(  
 \frac{\kappa^2}{2} -1 \right) p_L^{S,{\rm in}} - \frac{\kappa^2}{2\sqrt{6}}
  \tilde p_L^{S,{\rm in}} .  
\end{eqnarray}
If we choose $\kappa=\sqrt{2}$, the input-output 
relations for the light are
\begin{eqnarray}
x^{C} &=& p_{A+}^{\rm in} 
 + \frac{1}{\sqrt{2}} x_L^{C,{\rm in}}  - \frac{1}{\sqrt{6}}
  \tilde p_M^{C,{\rm in}} ,  \nonumber \\
p^{C} &=& - x_{A+}^{\rm in} 
 - \frac{1}{\sqrt{2}} x_M^{C,{\rm in}}  - \frac{1}{\sqrt{6}}
  \tilde p_L^{C,{\rm in}} , \nonumber  \\
x^{S} &=& -x_{A-}^{\rm in} 
 + \frac{1}{\sqrt{2}} x_L^{S,{\rm in}}  - \frac{1}{\sqrt{6}}
  \tilde p_M^{S,{\rm in}} ,  \nonumber \\
p^{S} &=& - p_{A-}^{\rm in} 
 - \frac{1}{\sqrt{2}} x_M^{S,{\rm in}} 
  - \frac{1}{\sqrt{6}} \tilde p_L^{S,{\rm in}} ,  
\end{eqnarray}
which is completely analogous to the single-cell results in Eq. (\ref{finaloutput}).
In this case the atomic sum- and difference-quadratures are written into
the cosine and sine sidebands of one of the output beams.


\section{Losses at the cell walls} 

\subsection{Vacuum input modes}
 
So far we have assumed perfect light transmission through the cell walls. In
reality, passing through the boundaries between different media causes losses
which result in decreasing the read-out fidelity. To estimate the effect of
losses, each boundary  is modeled by a beam-splitter which transforms 
the input light quadratures as $x^{\rm out}
= \sqrt{1-A}x^{\rm in}+\sqrt{A}x^{\rm vac}$, where $A$ is the effective loss
coefficient, $x^{\rm vac}$ is the  vacuum-noise field quadrature 
and analogous transformation applies for the $p$-quadrature. 

In this subsection we assume the input modes L and M in vacuum states.
We shall consider read-out with unity gain so that the mean values of
the quadratures are faithfully transferred from the atoms  onto light and the
readout fidelity of coherent states $|\alpha\rangle$ is constant and does 
not depend on $\alpha$.
The field attenuation $\sqrt{1-A}$ can be compensated by increasing the
coupling constant $\kappa$, but the extra vacuum noise decreases the quality 
of the read-out. It is also possible to counteract the losses by amplifying
the output light beam in a phase insensitive amplifier with gain
$G=2/[\kappa^2(1-A)]$. However, this also adds some thermal noise.

\begin{figure}[!t!]
\centerline{\epsfig{file=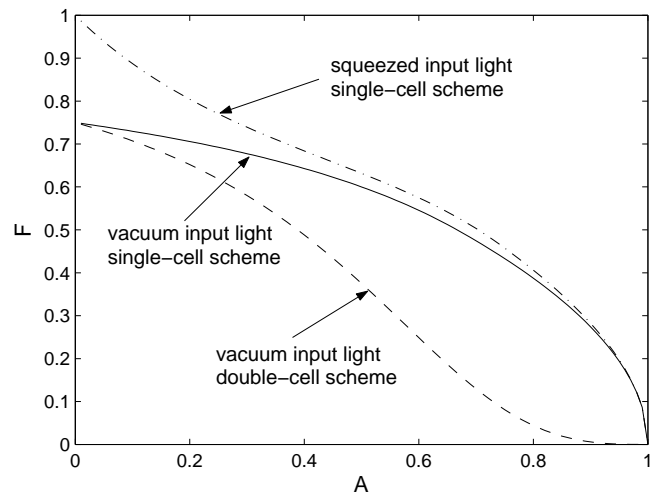,scale=0.7}}
\caption{Dependence of the fidelity of coherent state 
  read-out $\mathcal{F}$
  on the loss coefficient $A$ at the cell surface is shown for
  vacuum input light field and both the single-cell scheme (solid
  line) and the double-cell scheme (dashed line). The dot-dashed line
  depicts the fidelity achieved when using optimally squeezed input
  light in the single-cell scheme. 
\label{fig-absorpce}}
\end{figure}

After suitable modification of Eqs. (\ref{outquadratures})
and (\ref{output-kappa}) we find that for $A\leq 2/3$ 
the amplifier does not help and the best result is achieved with the
choice $\kappa=\sqrt{2/(1-A)}$ which leads to the read-out
transformation with added noise corresponding to mean 
thermal photon number
$\bar{n}$  and  coherent state read-out fidelity $\mathcal{F}$ given
by
\begin{eqnarray}
 \bar{n} = \frac{1}{3(1-A)}, \qquad \mathcal{F}=\frac{3(1-A)}{4-3A} .
\end{eqnarray}
Note that for small losses one gets the fidelity approximately
\begin{eqnarray}
  \mathcal{F} \approx \frac{3}{4} - \frac{3A}{16} .
\end{eqnarray}
 On the other hand, for $A>2/3$ we find that
the optimal coupling constant is $\kappa^2=2\sqrt{3/(1-A)}$, the
output signal needs to be amplified, and the resulting noise and
fidelity read,
\begin{equation}
\bar{n}=\frac{2}{\sqrt{3(1-A)}}-1, \qquad \mathcal{F}=\frac{1}{2}\sqrt{3(1-A)}.
\end{equation}
The dependence of $\mathcal{F}$ on $A$ can be seen in Fig. \ref{fig-absorpce}
(solid line). As can be seen, the scheme is rather robust with respect to
losses so that fidelity above $1/2$ is achieved with the loss coefficient 
up to $A=2/3$.

When using the two-cell scheme (Fig. \ref{fig-scheme}b), the influence of losses
is more complicated.
To make the discussion simple, let us consider the case when the mean values of
the quadratures are retrieved with unit gain without amplification of the output
beam and we also require that the cosine mode of the light beam does not couple
to the difference-quadrature atomic mode with quadratures $x_{A-}$
and $p_{A-}$. These conditions uniquely determine the
values of the coupling constants  which must be different in the two cells, 
in particular, we have
\begin{eqnarray}
 \kappa_1 = \sqrt{\frac{2}{(1-A)^3}}, \qquad
 \kappa_2 = \sqrt{\frac{2}{(1-A)}} .
\end{eqnarray}
The calculation of the read-out fidelity can be most easily carried out with the
help of the formalism of Gaussian completely positive maps 
\cite{Lindblad00,Eisert02,Sherson05b} which allows us to
describe the effects of losses in a particularly simple way. Assuming the 
atomic modes as well as the light beams are initially in coherent states, the
joint state of the atoms+light system remains Gaussian throughout the evolution
and can be fully characterized by the first and second moments of the 
quadrature operators. Moreover, since some quadratures remain uncoupled during
the readout procedure it suffices to consider only seven quadratures that can be
conveniently collected into a vector
$r=(x_L^C,p_M^C,p_L^S,\tilde{p}_L^S,\tilde{p}_M^C,p_{A}^{(1)},p_{A}^{(2)})$.
The covariance matrix which captures the quadrature fluctuations and correlations
is defined as follows, 
$\gamma_{jk}=\langle r_jr_k+r_kr_j\rangle-2\langle r_j\rangle\langle r_k\rangle$.
The initial covariance matrix is the identity matrix, 
$\gamma_{\mathrm{in}}=\openone$. After the passage through the first 
atomic ensemble, $\gamma$ transforms according to 
$\gamma_1=S_AS_1\gamma_{\mathrm{in}} S_1^TS_A^T+G$, where 
\[
S_1=\left(
\begin{array}{ccccccc}
1 & -\frac{\kappa_1^2}{4}& \frac{\kappa_1^2}{4}& \frac{\kappa_1^2}{4\sqrt{3}} 
& -\frac{\kappa_1^2}{4\sqrt{3}} & \frac{\kappa_1}{\sqrt{2}} & 0\\
0 & 1 & 0 & 0 & 0 & 0 & 0\\
0 & 0 & 1 & 0 & 0 & 0 & 0\\
0 & 0 & 0 & 1 & 0 & 0 & 0\\
0 & 0 & 0 & 0 & 1 & 0 & 0\\
0 & -\frac{\kappa_1}{\sqrt{2}} & \frac{\kappa_1}{\sqrt{2}} & 0 & 0 & 1 & 0\\
0 & 0 & 0 & 0 & 0 & 0 & 1
\end{array}
\right)
\]
describes the coupling of light with the atoms and the matrices 
$S_A=\sqrt{1-A}\openone_5\oplus\openone_2$ and $G=A\openone_5\oplus 0_2$
account for the losses on the cell wall. Here $\openone_j$ denotes an identity
matrix with $j$ rows and columns and  $0_j$ stands for $j\times j$ square 
matrix with all elements equal to zero. After the passage through the second 
atomic sample the covariance matrix reads,
\[
\gamma_{2}= S_AS_2(S_A\gamma_1 S_A^T+G)S_2^TS_A^T+G.
\]
This formula accounts for the losses on the two walls of the second glass 
cell as well as the interaction with the second atomic ensemble,
\[
S_2=\left(
\begin{array}{ccccccc}
1 & -\frac{\kappa_2^2}{4}& -\frac{\kappa_2^2}{4}& -\frac{\kappa_2^2}{4\sqrt{3}} 
& -\frac{\kappa_2^2}{4\sqrt{3}} & 0 & \frac{\kappa_2}{\sqrt{2}} \\
0 & 1 & 0 & 0 & 0 & 0 & 0\\
0 & 0 & 1 & 0 & 0 & 0 & 0\\
0 & 0 & 0 & 1 & 0 & 0 & 0\\
0 & 0 & 0 & 0 & 1 & 0 & 0\\
0 & 0 & 0 & 0 & 0 & 1 & 0\\
0 & -\frac{\kappa_2}{\sqrt{2}} & -\frac{\kappa_2}{\sqrt{2}} & 0 & 0 & 0 & 1
\end{array}
\right).
\]
The final interference of the beams L and M on a balanced beam splitter can be
again described by a symplectic matrix but for our purposes it suffices to
evaluate the variance of the resulting quadrature operator $x^C$ which is
related to the thermal noise added to retrieved state via
$\bar{n}=\langle(\Delta x^{C})^2\rangle-1/2$. It holds that
\[
\bar{n}=\frac{1}{4}(\gamma_{2,11}+\gamma_{2,22}+ 2\gamma_{2,12})-\frac{1}{2}.
\]

After tedious but straightforward algebra  one finds the added thermal noise 
and read-out fidelity to be
\begin{eqnarray}
 \bar{n}=\frac{1-A+\frac{1}{2}A^2}{3(1-A)^3}, \quad 
 \mathcal{F}=\frac{3(1-A)^3}{4-10A + \frac{19}{2}A^2-3A^3} .
\end{eqnarray} 
Thus, for small losses one finds
\begin{eqnarray}
 \mathcal{F} \approx \frac{3}{4} - \frac{3A}{8} ,
\end{eqnarray} 
i.e., the influence of losses is twice as large as in the single-cell scheme.
Note that the required ratio between the two coupling constants,
$\kappa_2/\kappa_1=(1-A)$ is naturally achieved in practice because
$\kappa\propto \sqrt{N_L}$ and the intensity of the light
beam is attenuated by a factor of $(1-A)^2$ when crossing the 
two cell walls that separate the first and second atomic ensembles 
(c.f. Fig. 1). The fidelity as a function of $A$ is plotted 
in Fig. \ref{fig-absorpce} (dashed line). 
As can be seen, with large losses the fidelity decreases much faster in the
two-cell scheme than in the single-cell one.
 

\subsection{Squeezed auxiliary modes}

So far we have assumed that the unused input light modes were in
vacuum  states.  In principle, the fidelity of the transfer
can be improved by using light with squeezed quadratures $x_L$,
$\tilde{p}_L$, $x_M$, and $\tilde{p}_{M}$. Let us first consider the
probably experimentally simplest case where we squeeze the $x$ quadratures 
of all modes in both beams L and M, irrespective of their temporal profile.
 Assuming that pure squeezed vacuum with the minimum uncertainty is
used, we have
\begin{equation}
\langle(\Delta x_{L}^{\rm in})^2\rangle=
\langle(\Delta x_{M}^{\rm in})^2\rangle=\frac{1}{2}e^{-2r},
\end{equation}
\begin{equation}
\langle(\Delta \tilde{p}_{L}^{\rm in})^2\rangle=
\langle(\Delta \tilde{p}_{M}^{\rm in})^2\rangle=\frac{1}{2}e^{2r}.
\end{equation}
The fidelity reads
\begin{equation}
\mathcal{F}=\frac{12}{12+3e^{-2r}+e^{2r}}.
\end{equation}
 The optimal squeezing that maximizes the fidelity is $e^{2r}=\sqrt{3}$ and 
 \begin{equation}
 \mathcal{F}_{max}=\frac{6}{6+\sqrt{3}} \approx 0.776,
 \end{equation}
which is only very small improvement over the read-out with light beams in
vacuum states. 

Much better results can be expected  when squeezing is
very selectively distributed among the modes such that
$x_{L}$, $x_{M}$, $\tilde{p}_{L}$, and $\tilde{p}_{M}$ are squeezed. In such a
case the fidelity can approach unity with increasing squeezing, provided
losses are negligible. Presence of losses at the boundaries decreases fidelity 
(see Fig. \ref{fig-absorpce}). For the sake of simplicity let us consider the
single-cell scheme. Optimal read-out is achieved when the
quadratures $\tilde{p}_M$ and $\tilde{p}_L$ are very strongly squeezed, while
the optimal squeezing  of the modes $L$ and $M$ is finite for $0<A<1$ 
and the variances of the four quadratures can be expressed in terms of a single
variance $V$ as follows, 
$\langle (\Delta x_L)^2\rangle=\langle (\Delta x_M)^2\rangle=V$ and 
$\langle (\Delta p_L)^2\rangle=\langle (\Delta p_M)^2\rangle=1/(4V)$.
We find that for $A \leq
A_{\mathrm{th}}=\frac{1}{2}(3-\sqrt{7/3})$ it is optimal to choose 
\[
\kappa^2=\frac{2}{1-A}, \qquad V=\frac{A}{2(1-A)},
\]
while for $A>A_{\mathrm{th}}$ we get 
\[
\kappa^2=2\sqrt{\frac{3(2-A)}{1-A}}, \qquad 
V=\frac{1}{4}\left|\kappa^2-2\right|.
\]
In this latter case, it is necessary to amplify the signal in the
output light mode in a phase insensitive amplifier with a gain
$G=2/[\kappa^2(1-A)]$. The resulting optimal fidelity can be obtained
from Eq. (\ref{Fidelitynbar}) where the mean number of noise photons reads
\begin{equation}
\bar{n}=\left\{
\begin{array}{ll}
A(1-A)+\frac{A}{3(1-A)}, & A \leq A_{\mathrm{th}}, \\[2mm]
\displaystyle
2A\sqrt{\frac{2-A}{3(1-A)}}-A, & A > A_{\mathrm{th}}.
\end{array}
\right.
\end{equation}
One finds that for $A\ll 1$ the fidelity decreases 
with the loss coefficient as $\mathcal{F} \approx 1-(4/3)A$. 

\section{Conclusion}

The proposed scheme for transfer of the quantum information from atomic samples
into the light pulses can be very efficient since only a single pass of the
beam through the medium is needed. Thus, no storage of the long (300 km) pulse
is necessary and the losses at the media boundary are minimized.  The price for
this operational simplicity is some  imperfectness of the read-out if vacuum
input fields are used: providing no losses occur, the  read-out fidelity  of
coherent states can approach the limiting value 3/4 which is still
substantially higher than the measure-and-prepare limit of 1/2. Unit fidelity
can be approached in our scheme if strongly squeezed fields in the relevant
input modes are used.  We have studied the influence of losses at the cell
walls, showing that the fidelity decreases with losses twice as fast when a
two-cell scheme is used in comparison to a single-cell scheme.  This also
confirms our motivation to search for simple, compact schemes for quantum
information exchange between field and material carriers. Our results can lead
to building quantum memory devices suitable for storing and straightforward
retrieving of information encoded in continuous variables.

\begin{acknowledgments}
This work was supported by the EU under projects COVAQIAL, 
QAP and QUACS, by the Czech Ministry of Education under 
the research project Measurement and Information in Optics (MSM6198959213), 
and by GA\v{C}R (202/05/0486).
\end{acknowledgments}


\end{document}